\documentclass[prb,aps,twocolumn,floatfix,superscriptaddress]{revtex4-2}
\usepackage{color}
\usepackage{amsmath}
\usepackage{latexsym}
\usepackage{amssymb}
\usepackage{amsmath}
\usepackage{braket}
\usepackage{amsfonts}
\usepackage{mathrsfs}
\usepackage{graphics,epstopdf}
\usepackage[colorlinks=true, citecolor=blue, urlcolor=blue, linkcolor=blue ]{hyperref}
\usepackage{epsf,graphics,graphicx}

\date{\today}
\usepackage{bigints}
\usepackage{multirow}
\usepackage{makecell}
\begin{document}

\title
  {Tailoring Phonon Transport in Bi-based Quaternary Half-Heusler alloys for Thermoelectric Applications}
  
\author{Supriya Ghosal}
\thanks{These authors contributed equally to this work.}
\author{Sayan Paul}
\thanks{These authors contributed equally to this work.}
\email{sayanpaul@jncasr.ac.in}
\author{Swapan K. Pati}
\email{pati@jncasr.ac.in}
\affiliation
{Theoretical Sciences Unit, School of Advanced Materials (SAMat), Jawaharlal Nehru Centre for Advanced Scientific Research, Bangalore 560064, India}
\date{\today}


\begin{abstract}
Quaternary half-Heuslers (QHHs) are substitutionally disordered derivatives of conventional half-Heusler (HH) compounds, which offers a potential route to modify phonon transport and enhance thermoelectric performance. Here, we investigate the thermoelectric properties of the quaternary half-Heuslers \(\mathrm{ZrHfRh_2Bi_2}\), \(\mathrm{ZrHfIr_2Bi_2}\), \(\mathrm{Zr_2RhIrBi_2}\), and \(\mathrm{Hf_2RhIrBi_2}\) using first-principles calculations and anharmonic lattice dynamics. Among four compounds, \(\mathrm{ZrHfRh_2Bi_2}\) and \(\mathrm{Zr_2RhIrBi_2}\) exhibit semiconducting electronic structure and others are semimetallic in nature. Both the  semiconducting compounds show favorable electronic transport properties. However, quaternary substitution substantially modifies the phonon dispersion, lowering the average phonon group velocity by about 17–22\% relative to the parent compounds while leaving the anharmonic scattering rates comparable. This suppresses the population (particle-like) thermal conductivity, partly offset by a small coherence contribution, and reduces $\kappa_l$ by about one-third. The combination of high power factor and low $\kappa_l$ results in high thermoelectric performance, with maximum \(ZT\) values of 2.28 and 2.18 at 1000 K for p-type \(\mathrm{ZrHfRh_2Bi_2}\) and \(\mathrm{Zr_2RhIrBi_2}\), respectively, with maximum thermoelectric power conversion efficiency $\sim$16.3\% for p-type ZrHfRh$_2$Bi$_2$. We further incorporate grain-boundary scattering to examine the effect of finite grain size on thermoelectric performance. Grain-boundary scattering suppresses both the charge-carrier mobility and $\kappa_l$, however the reduction of \(\kappa_l\) is much stronger. Thus, reducing the grain size enhances the thermoelectric performance, with maximum \(ZT\) values of 2.51 and 2.58 at 1000 K for p-type \(\mathrm{ZrHfRh_2Bi_2}\) and \(\mathrm{Zr_2RhIrBi_2}\), respectively, at a grain size of 30 nm. Our results demonstrate that combining substitutional disorder with grain-boundary engineering, provides an effective strategy to enhance the thermoelectric performance of HH compounds. 
\end{abstract}

\maketitle

\section{Introduction}
A major share of global energy consumption still originates from the combustion of fossil fuels, which causes significant environmental degradation, bio-diversity destruction and adverse health effects. This challenge has intensified the search for clean and renewable energy technologies. Thermoelectric materials offer a sustainable solid-state technology for converting the waste heat into electricity directly, therefore improving overall energy efficiency \cite{shi2020advanced, he2017advances, snyder2008complex}. The performance of a thermoelectric material can be quantified by a dimensionless quantity, called thermoelectric figure-of-merit ($ZT=\frac {S^2\sigma}{\kappa}T$), where $S$ denotes the Seebeck coefficient, $\sigma$ the electrical conductivity, $T$ the temperature and $\kappa=\kappa_l+\kappa_e$ the total thermal conductivity, composed of lattice thermal conductivity ($\kappa_l$) and electronic thermal conductivity ($\kappa_e$) \cite{shi2020advanced}. Therefore, achieving high thermoelectric performance requires the simultaneous enhancement of the power factor ($S^2\sigma$) and the suppression of thermal conductivity ($\kappa$). Several strategies have been devoted throughout the decades to enhance $ZT$ of thermoelectric materials, by decoupling the interdependent relationship between electronic and thermal transport properties in TE materials such as SnSe \cite{zhao2016snse, zhou2021polycrystalline}, AgSbTe$_2$ \cite{roychowdhury2021enhanced, zhang2025realizing}, Cu$_2$Se \cite{choo2024geometric, hu2024highly}, PbTe\cite{heremans2008enhancement} etc. 

In recent years, half Heusler compounds have emerged as a prominent class of thermoelectric materials owing to their excellent thermal stability, mechanical robustness, nontoxic composition, and intrinsically high power factors \cite{fu2015realizing, li2024half}. For instance, p-type (Nb$_{0.6}$Ta$_{0.4}$)$_{0.8}$Ti$_{0.2}$FeSb exhibit a maximum $ZT$ of $\sim$ 1.60 at 1200 K \cite{yu2018unique} and n-type  Ti$_{0.57}$Zr$_{0.4}$Al$_{0.02}$Ta$_{0.01}$NiSn$_{0.98}$Sb$_{0.02}$ achieves a peak $ZT$ of $\sim$ 1.39 at 870 K \cite{zhang2023ultralow}. Among the various half Heusler families, it is observed that \textit{X}FeSb (\textit{X} = V, Nb, and Ta) and  \textit{M}NiSn (\textit{M} = Ti, Zr, and Hf)-based systems demonstrate superior thermoelectric performances in the p and n-type regions respectively \cite{paul2025strategies}. In 2015, Gautier and co-workers introduced Bi-based half-Heusler compounds, opening a new compositional space for thermoelectric optimization \cite{gautier2015prediction}. Subsequent work by Zhu et al. revealed that Sn-doped ZrCoBi exhibits one of the highest reported $ZT$ values among half-Heusler materials, reaching $\sim$ 1.42 at 973 K \cite{zhu2018discovery}. Following this, in our previous work we carried out a systematic first-principles investigation of nine Bi-based HH compounds \textit{XY}Bi (\textit{X} = Ti, Zr, Hf and \textit{Y} = Co, Rh, Ir) to assess their thermoelectric applications \cite{paul2024enhanced}. Our study revealed excellent thermoelectric performance across this family, with n-type ZrRhBi and p-type HfRhBi exhibiting peak $ZT$ values of 1.98 and 1.93, respectively, owing to their high power factors. Nevertheless, a major disadvantage of HH compounds remains their high $\kappa_l$ compared to state-of-the-art good thermoelectric materials, which limits their thermoelectric performances \cite{paul2025strategies}. Thus, reducing $\kappa_l$ becomes one of the primary approach to enhance the thermoelectric performance of HH compounds. 

Various strategies have been used to suppress the $\kappa_l$ in different materials, including isoelectronic and aliovalent alloying\cite{liu2019design, han2023strong, zhu2018discovery}, microstructural control\cite{he2021high, yan2011enhanced, biswas2012high}, entropy engineering\cite{ghosh2024high, bhui2025atomic}, ferroelectric instability\cite{banik2019engineering, zhang2011anomalous}, and dimensionality reduction\cite{huang2019significant} etc. Among these approaches, alloying can simultaneously modify the electronic band structure and introduce additional point-defect scattering to optimize both electrical and thermal transport properties\cite{zhu2017compromise}. By introducing mass fluctuations and strain-field effects, substitutional alloying effectively scatters short-wavelength phonons at the atomic scale. Conversely, grain boundary scattering at the nanoscale and mesoscale can be enhanced by controlling the microstructure of the compound, thereby suppressing medium- to high-wavelength phonons \cite{biswas2012high}. Anand et al. proposed quaternary double (\textit{$XX'Y_2Z_2$}), triple (\textit{$XX'_2Y_3Z_3$}) and quadruple (\textit{$XX'_3Y_4Z_4$}) HH compounds, where \textit{X} and \textit{X'} are aliovalent elements maintaining the net zero valance balance \cite{anand2019double}. These quaternary HH compounds exhibit significantly lower $\kappa_l$ compared to the ternary counterparts due to the presence of alloy scattering mechanism associated with the disorder of two ternary sublattices \cite{sahni2024double, guo2020prediction, he2022improving}. Similarly, Isoelectronic quaternary HH compounds also effectively suppress $\kappa_l$ while exerting minimal influence on electrical transport, as they do not introduce charge disorder into the crystal lattice. For example, Zhou et al. reported a $\kappa_l$ of $\sim$ 18.5 Wm$^{-1}$K$^{-1}$ for TiCoSb at 300 K, which is markedly reduced to 6.4 and 6.9 Wm$^{-1}$K$^{-1}$ in  TiCo$_{0.5}$Rh$_{0.5}$Sb and Ti$_{0.5}$Zr$_{0.5}$CoSb respectively \cite{zhou2005disorder}. 

Motivated by this, we take quaternary half-Heusler (QHH) approach to reduce $\kappa_l$ and enhance the thermoelectric performance of our previously studied Bi-based HH compounds. We considered four Bi-based QHH compounds ZrHfRh$_2$Bi$_2$, ZrHfIr$_2$Bi$_2$, Zr$_2$RhIrBi$_2$, Hf$_2$RhIrBi$_2$. Among these four compounds, ZrHfIr$_2$Bi$_2$ and Hf$_2$RhIrBi$_2$ show semimetallic characteristics, so they are discarded for thermoelectric properties calculations. $\kappa_l$ of these QHH compounds reduced to 7-8 Wm$^{-1}$K$^{-1}$ at 300 K, which is attributed mainly to their lower phonon group velocity ($v_g$), while the phonon scattering rates remain comparable to those of the parent HH compounds. Consequently, we observe higher $ZT$ in the QHH compounds with the highest values being 2.28 for p-type ZrHfRh$_2$Bi$_2$ and 2.18 for p-type Zr$_2$RhIrBi$_2$ at 1000 K with maximum thermoelectric power conversion efficiency $\sim$ 16.3\%. Later, incorporating grain boundary scattering further suppresses $\kappa_l$ of these compounds, which simultaneously increases their ZT values to 2.51 for p-type ZrHfRh$_2$Bi$_2$ and 2.58 for p-type Zr$_2$RhIrBi$_2$ at 1000 K.

\section{Computational Methodology}
We performed ﬁrst-principles density functional theory (DFT) calculations within the Vienna Ab initio Simulation Package (VASP)\cite{vasp,vasp2} using the pseudopotentials derived from projected augmented wave (PAW) method\cite{kresse1999ultrasoft, PAW} to describe the interactions between the core and valance electrons. The Perdew, Burke and Ernzerhof functional (PBE) within the generalized gradient approximation (GGA) is used for describing the exchange-correlation energies \cite{GGA}. We set the threshold for energy and Hellmann-Feynman force convergence to 10$^{-8}$ eV and 10$^{-3}$ eV/$\mathrm{\AA}$, respectively. The electronic wavefunctions were expanded using a plane wave basis set with a 600 eV kinetic energy cutoff, and a $\Gamma$-centered 8$\times$8$\times$4 mesh was employed for the Brillouin zone sampling. Since PBE exchange correlation underestimates the band gap, we calculated the band structures using HSE06 hybrid functional considering spin-orbit coupling (HSE06+SOC), in which the exchange-correlation functional is built by combining 25\% exact Hatree-Fock exchange and 75\% PBE \cite{HSE}. The tetragonal ordered structure and the cubic SQS have different Brillouin zones, so their band structures cannot be compared directly along a common path. We therefore unfolded the SQS bands into the Brillouin zone of the parent ZrRhBi half-Heusler ($X$–$\Gamma$–$L$–$W$–$X$) using easyunfold \cite{easy1,easy2}, which allows the band gap and the VBM/CBM character of the disordered structure to be compared with those of the ordered phase at the PBE level. The electron-phonon renormalization of the band gap was calculated with the one-shot Zacharias-Giustino (ZG) configuration method \cite{zacharias2016one} as implemented in VASP \cite{vasp}. Harmonic phonons at $\Gamma$ were obtained by finite differences in a 2$\times$2$\times$2 supercell (96 atoms), and one ZG-displaced structure was generated for each temperature ($0, 100, \dots, 1000$ K in steps of 100 K). The band gap of each displaced supercell was evaluated at the PBE level of theory, and the renormalization was defined as $\Delta E_g (T) = E_g^{'} (T) - E_g^{static}$, where $E_g^{'} (T)$ is the band gap of the ZG-displaced supercell at temperature T, which includes the effect of electron-phonon interactions (zero-point motion at T = 0 K and thermal vibrations at finite T), and $E_g^{static}$ is the band gap of the undistorted structure with atoms at their equilibrium positions. The zero-point renormalization (ZPR) given by the $\Delta E_g (0)$ value. The phonon dispersions of Zr$_2$RhIrBi$_2$ and ZrHfRh$_2$Bi$_2$ were calculated using the frozen phonon method as implemented in the Phonopy package \cite{phonopy1,phonopy2}. For each system, a 2$\times$2$\times$2 supercell (96 atoms) of the primitive unit cell (12 atoms) was constructed to generate the symmetry inequivalent displaced structures. After that, the static DFT calculation was performed on these configurations to extract the second order interatomic force constants (IFCs).

The electronic transport properties, including the electrical conductivity ($\sigma$), Seebeck coefficient ($S$), and electronic thermal conductivity ($\kappa_e$), are evaluated by solving the Boltzmann transport equation (BTE) within the Onsager transport formalism, as implemented in the AMSET code\cite{Amset}. To calculate the electron relaxation time ($\tau_e$), four scattering mechanisms are considered: acoustic deformation potential (ADP) scattering arising from electron--acoustic phonon interactions, polar optical phonon (POP) scattering due to electron--optical phonon interactions described by the Fr\"ohlich model, ionized impurity (IMP) scattering, and piezoelectric (PIE) scattering resulting from the non-centrosymmetric crystal structure of the QHH compounds. We also include the effect of free charge carrier screening\cite{zhang2022phonon}, which weakens the long-range Coulomb interactions associated with polar optical phonons, and piezoelectric fields and reduces the scattering rates of POP and PIE mechanisms. This leads to an enhancement in the overall carrier mobility. The scattering rate of an electron from initial state $\ket{n\mathbf k}$ to final state $\ket{m\mathbf k+\mathbf q}$ is calculated using Fermi's golden rule\cite{ashcroft_mermin} as
\begin{equation}
    \tilde{\tau}^{-1}_{n\mathbf{k} \rightarrow m\mathbf{k} + \mathbf{q}}
    = \frac{2\pi}{\hbar} \left| g_{nm}(\mathbf{k}, \mathbf{q}) \right|^{2}
    \delta\!\left( \varepsilon_{n\mathbf{k}} - \varepsilon_{m\mathbf{k} + \mathbf{q}} \right),
\end{equation}
where $\varepsilon_{n\mathbf{k}}$ is the energy of state $\ket{n\mathbf k}$, and $\left| g_{nm}(\mathbf{k}, \mathbf{q}) \right|$ is the matrix element for scattering from state $\ket{n\mathbf k}$ into state $\ket{m\mathbf k+\mathbf q}$
\begin{equation}
    g_{nm}(\mathbf{k}, \mathbf{q})=\braket {m\mathbf k+\mathbf q|\Delta_{\mathbf q}V|n\mathbf k} 
\end{equation}
with $\Delta_{\mathbf q}V$ being any kind of electronic perturbation, such as electron-phonon interaction, electron-defect interaction etc.

\begin{figure*}[t] 
  \centering
  \includegraphics[width=\textwidth]{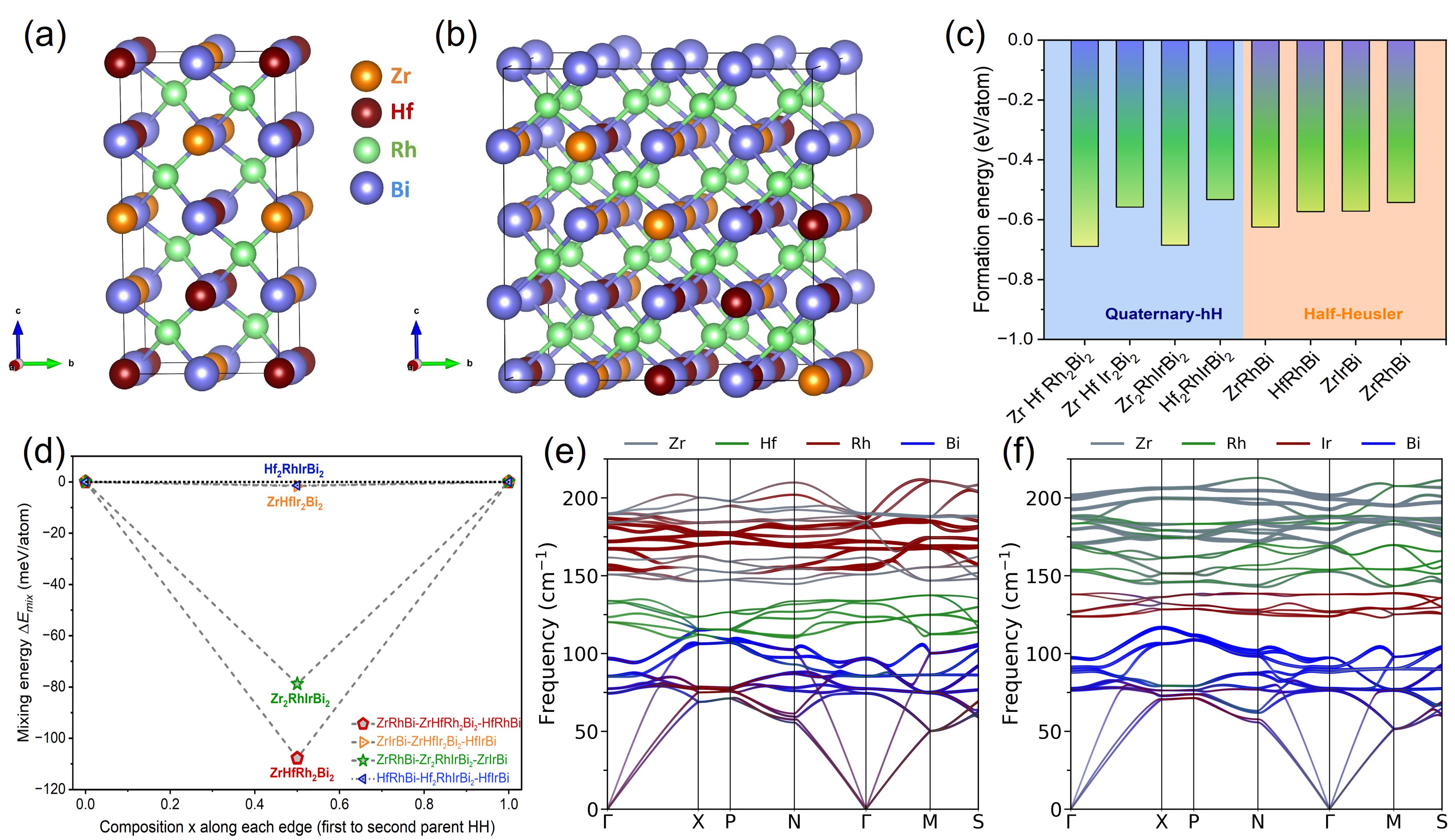}
  \caption{(a) Tetragonal and (b) disordered-cubic phases of ZrHfRh$_2$Bi$_2$. (c) Comparative formation energy plot of quaternary half-Heusler alloys and their half-Heusler counterparts. (d) Convex hull plot of tetragonal quaternary half-Heusler alloys as compared to their half-Heusler counterparts. Atom projected phonon dispersion spectra of (e) ZrHfRh$_2$Bi$_2$ and (f) Zr$_2$RhIrBi$_2$. }
  \label{fig:structure}
\end{figure*}

The lattice thermal conductivity ($\kappa_l$) is calculated based on Wigner formalism proposed by Simoncelli et al. \cite{simoncelli2019unified, simoncelli2022wigner}, implemented in the Phono3py package \cite{phono3py2,phonopy2}. In this formalism $\kappa_l$ is described by the equation
\begin{equation}
    \kappa_l^{\alpha\beta}=\kappa_p^{\alpha\beta}+\kappa_c^{\alpha\beta}
    \label{eq:kl}
\end{equation}
where $\kappa_p$ is referred to the ``populations thermal conductivity", which is the heat transport mechanism due to particle-like phonon propagation, where diagonal elements of the heat-flux matrix take part. $\kappa_p$ is obtained by solving linearized Boltzmann transport equation (LBTE) employing the harmonic and 3rd order interatomic force constants (IFCs) \cite{chaput2013direct}. To calculate 3rd order IFCs, a set of displaced configurations were generated for each compound from a 2$\times$2$\times$1 supercell of the primitive unit cell and 6 $\mathrm{\AA}$ cut-off radius. Then $\kappa_p$, along with the anharmonic phonon scattering rates ($\Gamma(q)$) and relaxation time ($\tau_{\mathrm{ph}}$) were calculated by solving LBTE using a 16$\times$16$\times$16 $q$-mesh. Conversely, $\kappa_c$ is reffered as ``coherences thermal conductivity", which arises from the off-diagonal terms of the heat-flux matrix. $\kappa_c$ is calculated from the equation
\small
\begin{eqnarray}
    \kappa_c^{\alpha \beta} = \frac{\hbar^2}{k_B T^2} \frac{1}{\mathcal{V}N_c} \sum_q \sum_{s \neq s'} \frac{\omega(q)_s + \omega(q)_{s'}}{2} v^\alpha(q)_{s,s'} v^\beta(q)_{s',s} \\ \nonumber  
    \times \frac{\omega(q)_s N^T(q)_s [N^T(q)_s + 1] + \omega(q)_{s'} N^T(q)_{s'} [N^T(q)_{s'} + 1]}{4[\omega(q)_{s'} - \omega(q)_s]^2 + [\Gamma(q)_s + \Gamma(q)_{s'}]^2} \\ \nonumber
    \times [\Gamma(q)_s + \Gamma(q)_{s'}]
    \label{eq:kc}
\end{eqnarray}
Here $\mathcal{V}$ is the volume of primitive unit cell, $N_c$ is the number of phonon eigenvectors,  $\omega(q)_s$ is the frequency of the phonon mode with $q$ wave vector and $s$ branch index, $v^{\alpha}(q)$ is the phonon group velocity in direction $\alpha$, $N^T(q)_s$ is the Bose-Einstein distribution at temperature T, and $\Gamma(q)$ is the phonon scattering rate. The off-diagonal terms are generally negligible for simple crystalline solids and become the dominant contributor in the disordered solids.

\section{Results and Discussion}

\subsection{Crystal Structure and Stability}

Quaternary half-Heuslers (QHH) are substitutionally disordered derivatives of conventional half-Heusler (HH) systems, in which one crystallographic sublattice is occupied by two distinct atomic species. The parent HH structure (\textit{XYZ}) crystallizes in a cubic lattice with space group $\mathrm{F\bar 43m}$, where \textit{X}, \textit{Y}, and \textit{Z} atoms occupy 4$a$ (0,0,0), 4$c$ (1/4, 1/4, 1/4) and 4$b$ (1/2, 1/2, 1/2) Wyckoff positions, respectively. In QHH compounds, one of the lattice sites (\textit{X}, \textit{Y} or \textit{Z}) is contributed by two atoms and forms $XX'Y_2Z_2$, $X_2YY'Z_2$ or $X_2Y_2ZZ'$. Here $X$, $X'$ or $Y$, $Y'$ or $Z$, $Z'$ can be isoelectronic or aliovalent elements. FIG. \textcolor{blue}{S1, Supplemental Material} shows the structure of QHH compound ZrHfRh$_2$Bi$_2$, in which Zr and Hf have equal occupancy at $X$-site (half orange and half maroon colour), Rh at $Y$ site (green colour) and Bi at $Z$ site (blue colour). 

\begin{table*}
    \centering
    \setlength{\tabcolsep}{16pt}
    \caption{Theoretically optimized lattice constants, relative energy difference ($\Delta E$), and formation energies ($ E_f$ ) of tetragonal and SQS structures of four QHH compounds.}
    \label{tab:structure}
    \begin{tabular}{llccc}
        \hline\hline
        Compound & Crystal & Optimized lattice & $\Delta E$ & $E_f$ \\
        & structure & parameters (\AA) & (meV/atom) & (meV/atom) \\
        \hline
        \multirow{2}{*}{ZrHfRh$_2$Bi$_2$} 
            & Tetragonal & $a = b =  6.443, c = 12.885$ & 0.0 & $-688.79$ \\
            & Cubic (SQS)        & $a = b = c = 12.885$ & 0.17 & $-688.61$ \\
        \multirow{2}{*}{ZrHfIr$_2$Bi$_2$} 
            & Tetragonal & $a = b = 6.469, c = 12.937$ & 0.0 & $-557.82$ \\
            & Cubic (SQS)        & $a = b = c = 12.938$ & 0.20 & $-557.62$ \\
        \multirow{2}{*}{Zr$_2$RhIrBi$_2$} 
            & Tetragonal & $a = b = 6.477, c = 12.952$ & 0.0 & $-684.82$ \\
            & Cubic (SQS)        & $a = b = c = 12.885$ & 74.94 & $-609.88$ \\
        \multirow{2}{*}{Hf$_2$RhIrBi$_2$} 
            & Tetragonal & $a = b = 6.442, c = 12.876$ & 0.0 & $-532.53$ \\
            & Cubic (SQS)        & $a = b = c = 12.716$ & 85.99 & $-446.53$ \\
        \hline\hline
    \end{tabular}
\end{table*}
To model the disordered configuration of FIG. \textcolor{blue}{S1, Supplemental Material}, two distinct structural models were considered. The first one is tetragonal ordered structure, analogous to the ordered quaternary HH structure, reported in ref \cite{anand2019double} (FIG.\ref{fig:structure}b). The space group of this structure is $\mathrm{I\bar 42d}$ with 24 atoms in the conventional unit cell. The second one is a cubic special quasirandom structure (SQS)(FIG.\ref{fig:structure}c) generated with the ATAT code, which mimics the statistical distribution of atomic disorder on the mixed sublattice. For the SQS structure, we have taken a supercell of 96 atoms, maintaining cubic symmetry by implementing the cutoff radii 8 {\AA}, 7 {\AA} and 6 {\AA} for pair, three-body and four-body correlation, respectively. SQS is useful to treat random solid solutions at the DFT level in a way that replicates the pair and multibody correlations found in a completely random lattice, whereas the tetragonal structure only captures complete site ordering. The optimized lattice parameters of all structures are mentioned in TABLE \ref{tab:structure}. We calculate the formation energies of both structures using the equation,  
\begin{equation}
    E_f=\frac{1}{N}\left(E_t-\sum_{i=1}^N n_i\mu_i\right),
    \label{eq:formation}
\end{equation}
where $E_t$ is the total energy of the system, $n_i$ and $\mu_i$ are the number of atoms and chemical potential corresponding to species $i$, and $N$ is the total number of atoms in the unit cell. The chemical potential ($\mu_i$) for each element is calculated from the total energy of experimentally reported bulk structures. FIG.\ref{fig:structure}c compares the formation energies of the QHH compounds with those of their parent HH counterparts. All four QHH compounds have large negative formation energies, comparable to or lower than those of ZrRhBi, HfRhBi, ZrIrBi, and HfIrBi. To assess their stability against decomposition into the parent HH phases, we calculated the mixing energy,
\begin{equation}
    \Delta E_{mix} = \frac{1}{N}\left[E_{QHH} - 2(E_{HH_1} + E_{HH_2})\right]
\end{equation}
where HH$_1$ and HH$_2$ are the two parent half-Heuslers (e.g., ZrRhBi and HfRhBi for ZrHfRh$_2$Bi$_2$) and $E_{QHH/HH}$ represent corresponding total energy. The factor 2 comes from the formula unit mismatch as primitive QHH formula unit contains 12 atoms. As shown in FIG.\ref{fig:structure}d, $\Delta E_{mix}$ is negative for ZrHfRh$_2$Bi$_2$ ($\approx$ -107 meV/atom) and Zr$_2$RhIrBi$_2$ ($\approx$ -78 meV/atom), so these two compounds are stable against phase separation into their parent HH constituents. In contrast, $\Delta E_{mix}$ of ZrHfIr$_2$Bi$_2$ and Hf$_2$RhIrBi$_2$ are close to zero, indicating that they are only marginally stable with respect to the parent HH phases. TABLE \ref{tab:structure} shows that both the tetragonal ordered and cubic disordered structures possess negative $E_f$, which verifies their energetic stability. The ordered tetragonal structures are lower in energy than the corresponding SQS structures for all four compounds (TABLE \ref{tab:structure}). In alloys, the competition between ordered and disordered arrangements is governed by the free energy $\Delta G = \Delta E - T \Delta S_{\text{conf}}$, so disordered phases become increasingly favoured at elevated temperatures through configurational entropy \cite{zunger1990special,anand2019double,ghosalth}. For Zr/Hf mixing on the X site, the cubic SQS and tetragonal structures have nearly identical formation energies (0.2 meV/atom difference), so the SQS is expected to be stabilized by configurational entropy at high temperatures, whereas at low temperatures the tetragonal structure serves as a good model of the disordered phase. For Rh/Ir mixing on the Y site (Zr$_2$RhIrBi$_2$ and Hf$_2$RhIrBi$_2$), however, the tetragonal phase is favoured by 75-86 meV/atom, which configurational entropy (about 20 meV/atom at 1000 K) cannot overcome. We therefore use the tetragonal phases of all QHH compounds throughout this work, thereby reducing computational cost. These tetragonal phases of QHH alloys exhibit no imaginary frequency phonon modes in their phonon dispersion, verifying their dynamical stability [FIG.\ref{fig:structure}e, \ref{fig:structure}f, and \textcolor{blue}{S11, Supplemental Material}]. To ensure mechanical stability, QHH alloys must satisfy Born-Huang's criteria for materials having tetragonal symmetry as follows \cite{mouhat2014necessary}.
\begin{equation}
\begin{split}
    C_{11} > |C_{12}|,~& 2C_{13}^2 < C_{33}(C_{11} + C_{12}),\\ &C_{44} > 0,~ C_{66} > 0
\end{split}
\end{equation}
Here, $C_{ij}$'s are elastic coefficients. Elastic coefficients for QHH alloys are tabulated in TABLE \textcolor{blue}{S1, Supplemental Material}, clearly satisfy Born-Huang's criteria, ensuring mechanical stability. Therefore, by satisfying energetic, dynamical and mechanical stability criteria, QHH alloys are considered and well as feasible for experimental realization.  

\begin{figure}[t] 
  \centering
  \includegraphics[width=\columnwidth]{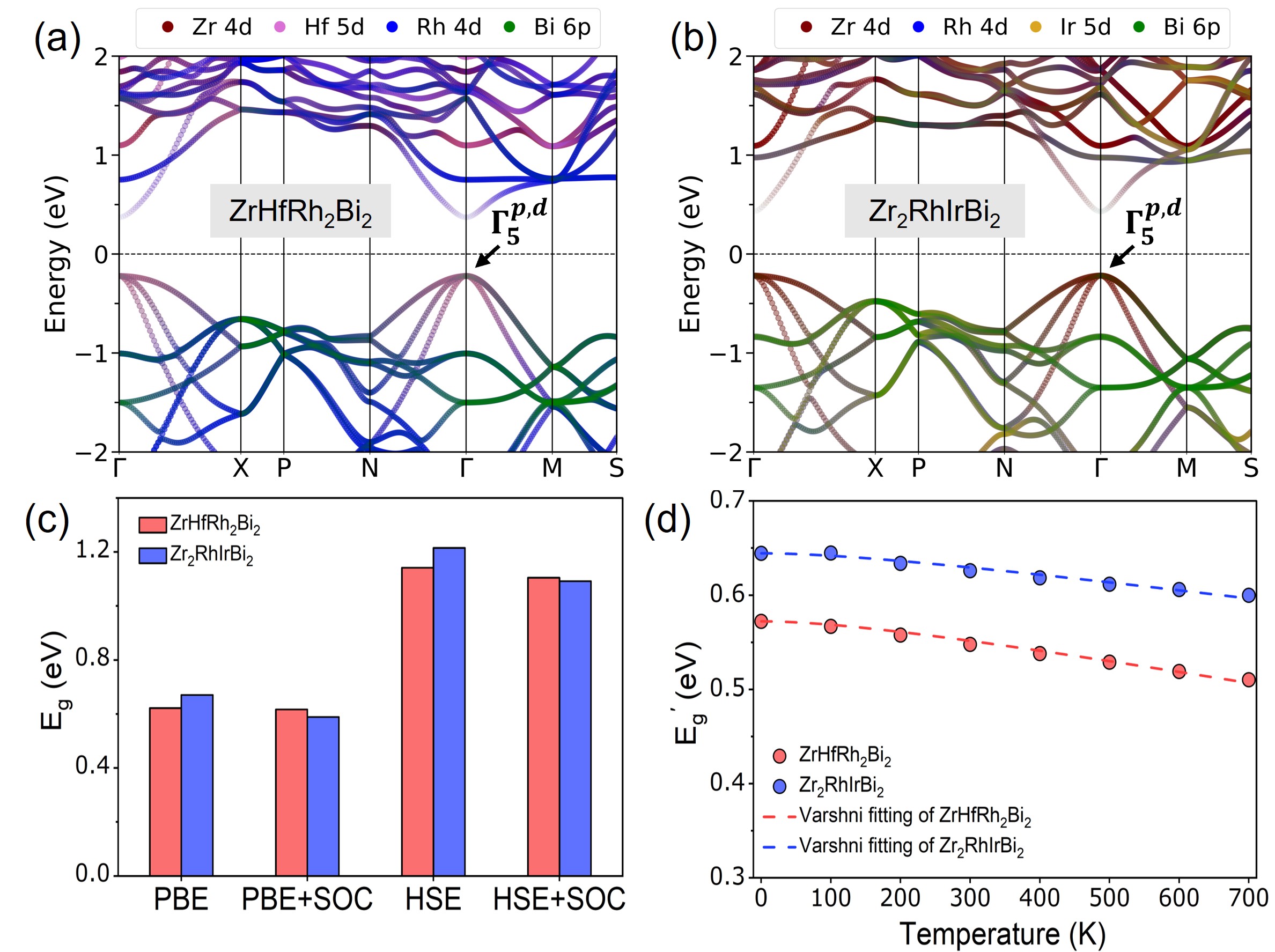}
  \caption{$p$- and $d$-Orbital-projected electronic band structures of (a) ZrHfRh$_2$Bi$_2$ and (b) Zr$_2$RhIrBi$_2$. (c) Electronic band gaps of ZrHfRh$_2$Bi$_2$ and Zr$_2$RhIrBi$_2$ using PBE and HSE06 functionals with and without including spin-orbit coupling (SOC) effect. (d) Temperature-dependent electronic band gaps $E_g^{'} (T)$ of ZrHfRh$_2$Bi$_2$ and Zr$_2$RhIrBi$_2$ based on ZG-displaced supercell structure, with the corresponding Varshni fits shown as red and blue dashed lines, respectively.}
  \label{fig:band}
\end{figure}

\subsection{Electronic Structure and Charge Carrier Transport}

The orbital-projected band structure calculated using PBE functional of ZrHfRh$_2$Bi$_2$ and Zr$_2$RhIrBi$_2$ are shown in FIG. \ref{fig:band}a and \ref{fig:band}b for $p$- and $d$-orbitals and in FIG. \textcolor{blue}{S2, Supplemental Material} for $s$-orbitals. Similarly, the orbital projected band structure of ZrHfIr$_2$Bi$_2$ and  Hf$_2$RhIrBi$_2$ are shown in FIG. \textcolor{blue}{S3, Supplemental Material}. Both ZrHfRh$_2$Bi$_2$ and Zr$_2$RhIrBi$_2$ are direct-band-gap semiconductors with PBE band gaps of 0.62 and 0.67 eV, respectively, whereas ZrHfIr$_2$Bi$_2$ and  Hf$_2$RhIrBi$_2$ exhibit semimetallic behavior. The calculated band gaps of ZrHfRh$_2$Bi$_2$ and Zr$_2$RhIrBi$_2$ obtained using the PBE, PBE+SOC, HSE06, and HSE06+SOC functionals are summarized in Fig.~\ref{fig:band}c. The bonding of these QHH compounds closely resembles that of conventional HH compounds. For ZrHfRh$_2$Bi$_2$ and Zr$_2$RhIrBi$_2$ the VBM is contributed by $X/X'$-$d$ and $Z$-$p$ orbitals and is characterized by $\Gamma^{p,d}_5$ irreducible representation (FIG. \ref{fig:band}a and \ref{fig:band}b). In contrast, CBM is dominated by Bi-$s$ orbital, which is characterized by $\Gamma^{s}_1$ irreducible representation (FIG. \textcolor{blue}{S2, Supplemental Material}). In ZrHfIr$_2$Bi$_2$ and  Hf$_2$RhIrBi$_2$, the $\Gamma^{s}_1$ state drops below the $\Gamma^{p,d}_5$ state, resulting in band inversion and the emergence of a semimetallic electronic structure. The observed band inversion is analogous to that reported for HfIrBi and suggests the possibility of topological phase transitions under compressive stress, as shown by Wang et al. \cite{wang2016topological}. We discarded the semimetallic compounds ZrHfIr$_2$Bi$_2$ and  Hf$_2$RhIrBi$_2$ for further thermoelectric properties calculations due to the limitations of our computational methodology. Since the tetragonal and SQS structures of ZrHfRh$_2$Bi$_2$ differ in energy by only 0.2 meV/atom, we also examined whether the two models give the same electronic band structure. The SQS bands, unfolded into the ZrRhBi Brillouin zone (FIG. \textcolor{blue}{S4, Supplemental Material}), reproduce the band gap and the dominant VBM and CBM orbital contributions of the tetragonal phase, with only minor spectral broadening, which justifies the use of the tetragonal model in the transport calculations.

In both semiconducting compounds, the band gap decreases monotonically with increasing temperature due to electron-phonon interactions, as shown in Fig. \ref{fig:band}d. Within the PBE approximation, the static-lattice gap of ZrHfRh$_2$Bi$_2$ is 0.62 eV. Including zero-point vibrations reduces it by 49 meV (ZPR = -49 meV) to 0.572 eV at 0 K, and the gap further decreases to 0.54 eV at 300 K and 0.51 eV at 700 K. For Zr$_2$RhIrBi$_2$, the static gap of 0.67 eV is reduced by 26 meV to 0.64 eV at 0 K, and then to 0.62 eV at 300 K and 0.61 eV at 700 K. This bandgap narrowing arises from the electron-phonon coupling, which increases with the phonon occupation at higher temperatures. The temperature-dependent gaps, including the zero-point value, were fitted using the empirical Varshni relation,
\begin{equation}
E_g^{'}(T)=E_g^{'}(0)-\frac{\alpha T^2}{T+\beta},
\end{equation}
where $E_g^{'}(0)$ represents the band gap at 0 K, $\alpha$ indicates electron-phonon coupling strength, and $\beta$ is a material-dependent parameter, which is related to Debye temperature for some materials. The values of $\alpha$ and $\beta$ for ZrHfRh$_2$Bi$_2$ are $1.27\times10^{-4}$ eV/K and 253 K, respectively, and for Zr$_2$RhIrBi$_2$ are $9.42\times10^{-5}$ eV/K and 263 K, respectively. These $\beta$ values of ZrHfRh$_2$Bi$_2$ and Zr$_2$RhIrBi$_2$ are close to their Debye temperatures ($\theta_D$) as obtained from the mechanical properties (TABLE \textcolor{blue}{S1, Supplemental Material}). Though $E_g$ decreases with temperature, both compounds retain a semiconducting nature with a finite band gap throughout the investigated temperature range.

\begin{figure}[t] 
  \centering
  \includegraphics[width=\columnwidth]{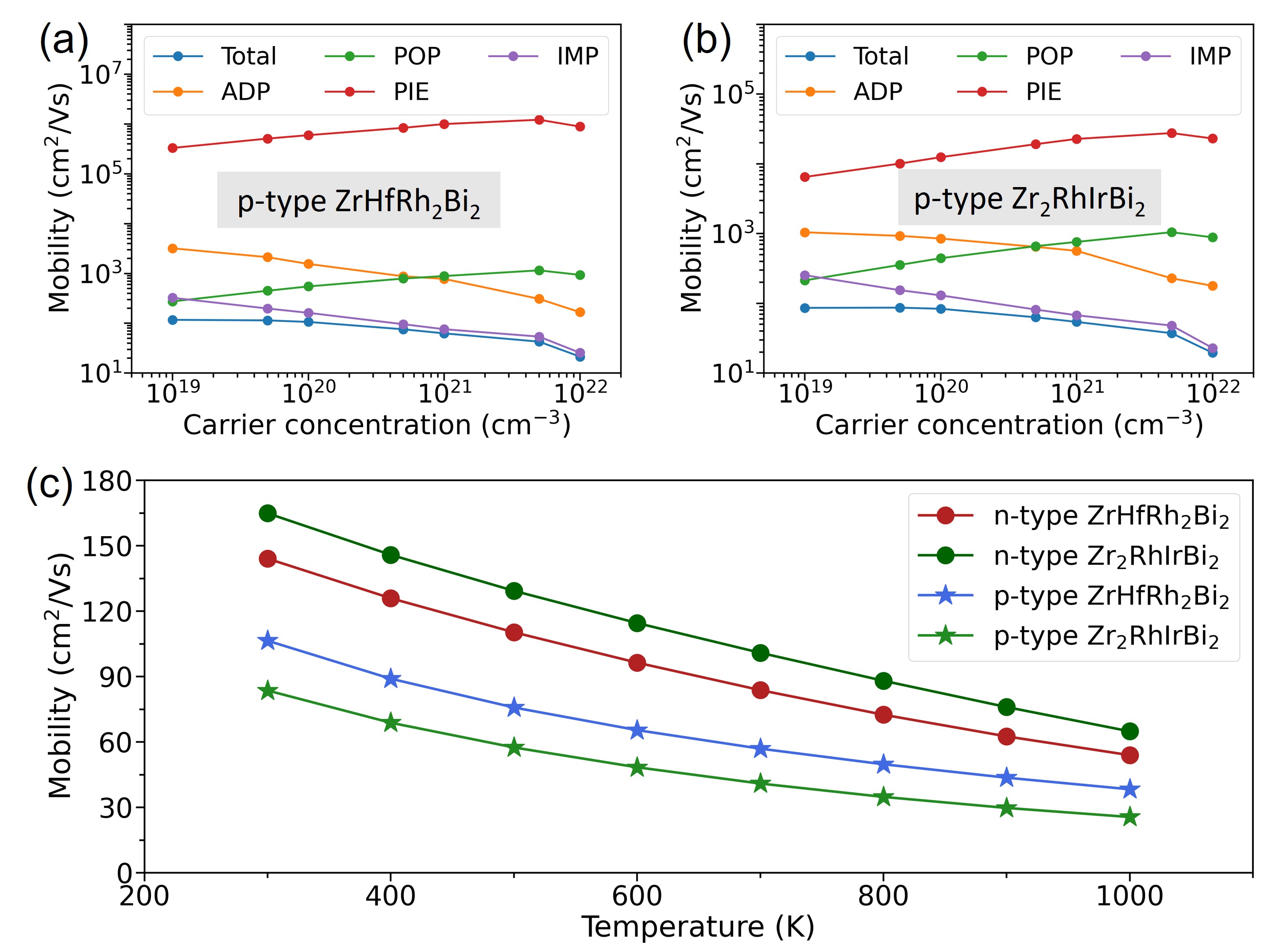}
  \caption{Contribution of different scattering mechanisms to the charge carrier mobility as a function of carrier concentration in (a) p-type ZrHfRh$_2$Bi$_2$ and (b) p-type Zr$_2$RhIrBi$_2$ at 300 K. (c) Temperature dependence of the total charge carrier mobility of ZrHfRh$_2$Bi$_2$ and Zr$_2$RhIrBi$_2$ for both n-type and p-type carriers at a carrier concentration of $1\times10^{20}$ cm$^{-3}$.}
  \label{fig:el_trans}
\end{figure}

We have calculated charge carrier mobility of ZrHfRh$_2$Bi$_2$ and Zr$_2$RhIrBi$_2$ in a temperature range 300--1000 K. FIG. \ref{fig:el_trans}a and \ref{fig:el_trans}b represent the contribution of different scattering mechanisms to the charge carrier mobility as a function of carrier concentration for p-type ZrHfRh$_2$Bi$_2$ and Zr$_2$RhIrBi$_2$, respectively, at 300 K. It is observed that in both systems, impurity scattering (IMP) is the primary mechanism to limit the total charge carrier mobility. At low carrier concentrations, polar optical phonon scattering (POP) is the next most important mechanism, stronger than acoustic deformation potential scattering (ADP). This is because the long-range electric field of LO phonons couples strongly to holes when free-carrier screening is weak. As the carrier concentration increases, the growing number of free carriers screens the long-range electric field and the matrix element in Eq. \textcolor{blue}{S12, Supplemental Material}  modifies from $\frac{1}{|\mathbf q+\mathbf G|}$ dependence to  $\frac{1}{|\mathbf q+\mathbf G|+\beta_{\infty}}$ dependence, where $\beta_{\infty}$ is the inverse screening length that depends on the temperature, carrier concentration, and high-frequency dielectric constant. This weakens the electron-LO-phonon interaction and suppresses POP scattering. ADP scattering, however, arises from short-range strain fluctuations and is largely unaffected by screening. As a result, ADP becomes relatively stronger and eventually overtakes POP at high carrier concentrations. Piezoelectric scattering (PIE) remains very weak throughout, indicating that the piezoelectric coupling is intrinsically weak compared to the deformation-potential and polar-optical interactions in these materials. FIG. \ref{fig:el_trans}c depicts the variation of charge carrier mobility of ZrHfRh$_2$Bi$_2$ and Zr$_2$RhIrBi$_2$ with respect to temperature for both n- and p-type regions at a carrier concentration of $1\times10^{20}$ cm$^{-3}$. We observe relatively lower charge-carrier mobility in the p-type region for both compounds. This lower mobility in the p-type region is likely associated with a higher hole effective mass, and more degenerate valence bands compared to the conduction bands, as shown in Table \ref{tab:effective_mass}. The decrease in mobility with increasing temperature is expected because the occupation number of active phonon modes grows with T (following Bose–Einstein statistics), which raises the electron–phonon scattering rates (ADP and POP) and lowers the mobility at the higher temperatures.

\begin{table}
\centering
\small
\renewcommand{\arraystretch}{1.5}
\setlength{\tabcolsep}{6pt}
    \begin{tabular}{c|c c c|c c c}
        \hline
        \hline
        \multicolumn{1}{c|}{Compound} & \multicolumn{3}{c|}{$\bm{m_h^*(m_0)}$$\bm{m_h^*(m_0)}$} & \multicolumn{3}{c}{$\bm{m_e^*(m_0)}$}\\ 
          & $\Gamma$-X & $\Gamma$-N & $\Gamma$-M & $\Gamma$-X & $\Gamma$-N & $\Gamma$-M \\ \hline
        ZrHfRh$_2$Bi$_2$ & 2.84  & 1.03  & 0.48  & 0.27  & 0.26  & 0.26   \\
         & 0.46  & 0.49 & 0.48 & --  & --  & --   \\
         & 0.13 & 0.14 & 0.17 & -- & -- & --  \\
        \hline
        Zr$_2$RhIrBi$_2$ & 3.05 & 1.22 & 0.52 & 0.18 & 0.18 & 0.18  \\
         & 0.54 & 0.53 & 0.53 & -- & -- & -- \\
         & 0.12 & 0.12 & 0.14 & -- & -- & --  \\
        \hline
        \hline
        \end{tabular}
    \caption{Band effective mass of hole ($m_h^*$) and electron ($m_e^*$) in different directions from VBM and CBM. The VBM is triply degenerate at $\Gamma$ high-symmetry point.}
    \label{tab:effective_mass}
\end{table}

\begin{figure}
    \centering
    \includegraphics[width=\columnwidth]{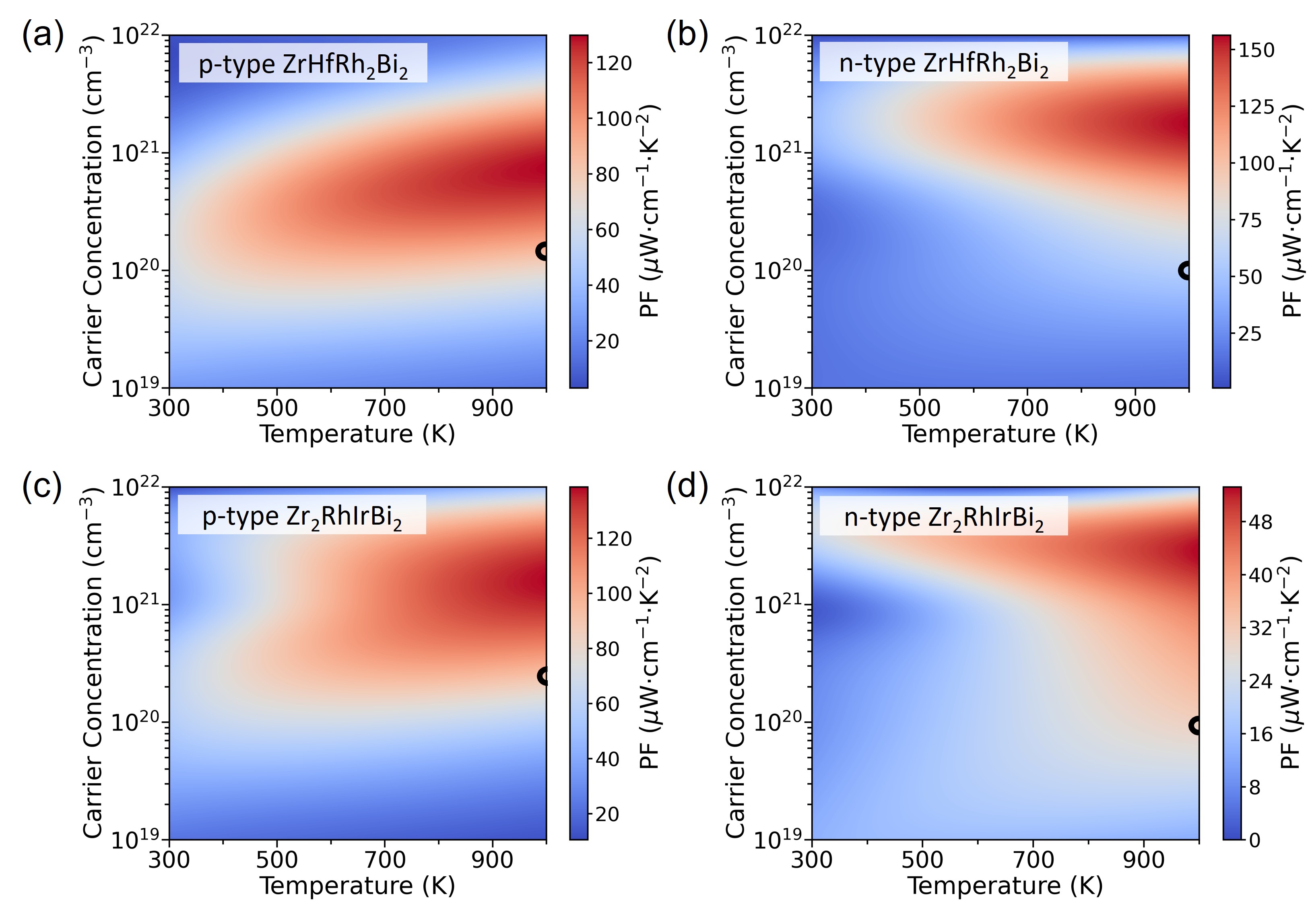}
    \caption{Thermoelectric power factor ($PF$) of (a) p-type ZrHfRh$_2$Bi$_2$, (b) n-type ZrHfRh$_2$Bi$_2$, (c) p-type Zr$_2$RhIrBi$_2$, and (d) n-type Zr$_2$RhIrBi$_2$ as functions of temperature and charge carrier concentration. The black circles indicate the carrier concentrations corresponding to the maximum $ZT$ in each compound.}
    \label{fig:power_factor}
\end{figure}

The calculated electrical conductivity ($\sigma$) and Seebeck coefficient ($S$) as functions of charge carrier concentration and temperature for n-type ad p-type ZrHfRh$_2$Bi$_2$ and Zr$_2$RhIrBi$_2$ are shown in FIG. \textcolor{blue}{S7, Supplemental Material}. Similarly, FIG. \ref{fig:power_factor} shows the colourmap of power factor (PF) with respect to carrier concentration and temperature. For all compounds, the highest PF values are obtained within the carrier concentration range of $1\times10^{20}$ to $5\times10^{21}$ cm$^{-3}$ and at 1000 K temperature. The black circles in FIG. \ref{fig:power_factor} indicate the carrier concentrations corresponding to the maximum $ZT$ for both n-type and p-type doping. Notably, the maximum $ZT$ does not coincide with the carrier concentration corresponding to the highest PF, because the electronic thermal conductivity ($\kappa_e$) increases significantly at higher carrier concentrations (FIG. \textcolor{blue}{S8, Supplemental Material}), which reduce the overall $ZT$ at those carrier concentrations.

\subsection{Phonon and Thermal Transport}

\begin{figure}[t] 
  \centering
  \includegraphics[width=\columnwidth]{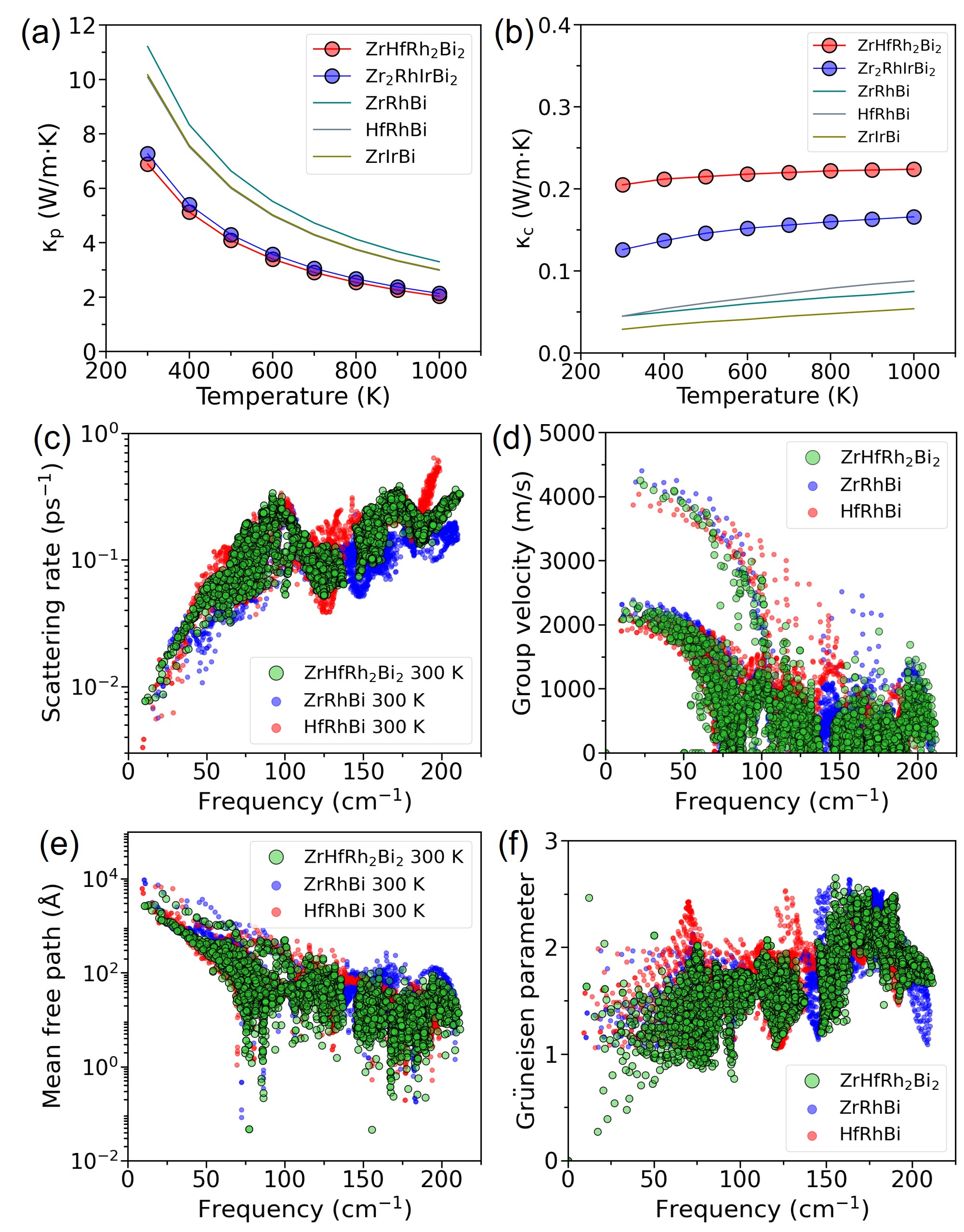}
  \caption{(a) Particle-like ($\kappa_p$) and (b) wave-like ($\kappa_c$) thermal conductivity of QHH compounds ZrHfRh$_2$Bi$_2$ and Zr$_2$RhIrBi$_2$ compared to their half-Heusler counterparts. Comparison of (c) phonon scattering rate at 300 K, (d) phonon group velocity ($v_g$), (e) mean free path at 300 K, and (f) mode Gr$\ddot{u}$neisen parameter ($\gamma$) between QHH compound ZrHfRh$_2$Bi$_2$ and corresponding HH compounds ZrRhBi, and HfRhBi.}
  \label{fig:kappa}
\end{figure}

We have calculated atom-projected phonon dispersion of ZrHfRh$_2$Bi$_2$ and Zr$_2$RhIrBi$_2$ as shown in FIG. \ref{fig:structure}e and \ref{fig:structure}f. The absence of imaginary phonon modes throughout the Brillouin zone confirms the dynamical stability of both compounds. As shown in FIG. \ref{fig:structure}e, the low-frequency optical phonon modes in ZrHfRh$_2$Bi$_2$ are predominantly contributed by Bi atoms, whereas the intermediate-frequency optical modes are contributed by Hf atoms and the high-frequency optical modes are primarily associated with the vibrations of Zr and Rh atoms. Similarly, for Zr$_2$RhIrBi$_2$, the low-frequency optical modes are dominated by Bi atoms, the intermediate-frequency optical modes by Ir atoms, and the high-frequency optical modes by Zr and Rh atoms, as shown in FIG. \ref{fig:structure}. This frequency distribution arises mainly from the atomic mass of particular atoms, with the heavier Bi atoms predominantly contributing to the low-frequency vibrations, while the lighter Zr and Rh atoms contribute mainly to the high-frequency optical modes. 

FIG. \ref{fig:kappa}a and \ref{fig:kappa}b compare the population ($\kappa_p$) and coherence ($\kappa_c$) contributions to the $\kappa_l$ of the QHH compounds and their corresponding HH counterparts. We observed that, while thermal transport in the ternary HH compounds is almost entirely governed by $\kappa_p$, the QHH compounds exhibit a substantial coherence contribution. However, $\kappa_p$ is significantly suppressed in QHH compounds, leading to a overall suppression of $\kappa_l$ as $\kappa_p$ has the major contribution to $\kappa_l$ (FIG. \textcolor{blue}{S12, Supplemental Material}). In HH compounds $\kappa_p$ varies 11.21--3.01 Wm$^{-1}$K$^{-1}$ at the temperature range 300--1000 K, and $\kappa_c$ is negligible, which varies 0.03--0.09 Wm$^{-1}$K$^{-1}$ at the same temperature range. In QHH compounds $\kappa_p$ reduced to 7.27--2.03 Wm$^{-1}$K$^{-1}$ and $\kappa_c$ increased to 0.20--0.22 Wm$^{-1}$K$^{-1}$ at 300--1000 K temperature range. This enhancement in $\kappa_c$ originates from the reduced average phonon interband spacing, which is calculated as $\Delta\omega_{\text{avg}}=\frac{\omega_{\text{max}}}{3\text{N}_{\text{at}}}$. As listed in Table \ref{tab:average}, $\Delta\omega_{\mathrm{avg}}$ decreases from $\sim$ 19--23 cm$^{-1}$ in the ternary HH compounds to about 5 cm$^{-1}$ in the QHH compounds. The reduced interband spacing increases the overlap of phonon linewidth between neighboring phonon branches, thereby facilitating coherent interband heat transport and enhancing the coherence contribution to $\kappa_l$.  

\begin{table}
    \centering
    \renewcommand{\arraystretch}{1.5}
    \begin{tabular}{lcccc}
        \hline
        \hline
        Compound & $\Delta\omega_{\text{avg}}$ (cm$^{-1}$)& $\bar{v_g}$ (m/s) & $\bar{\tau}$ (ps) & $\bar\gamma$  \\
        \hline
        ZrHfRh$_2$Bi$_2$  & 5.81 &  1223 & 20.26 & 1.73\\
        Zr$_2$RhIrBi$_2$  & 5.90 & 1151 & 26.42 & 1.76 \\
        ZrRhBi  & 23.27 & 1468 & 26.86 & 1.80\\
        ZrIrBi  & 23.20 & 1424 & 27.55 & 1.82\\
        HfRhBi  & 19.12 & 1507 & 21.69 & 1.80\\
        \hline
        \hline
    \end{tabular}
    \caption{Average phonon interband spacing ($\Delta\omega_{\text{avg}}$), average phonon group velocity ($\bar{v_g}$), and average phonon relaxation time ($\bar{\tau}$) of QHH compounds ZrHfRh$_2$Bi$_2$ and Zr$_2$RhIrBi$_2$, compared to their HH counterparts ZrRhBi, ZrIrBi, and HfRhBi.}
    \label{tab:average}
\end{table}

To elucidate the origin of the suppressed population contribution, we further analyzed the phonon scattering rate and phonon group velocity ($v_g$) as functions of phonon frequency. FIG. \ref{fig:kappa}c and \ref{fig:kappa}d compare these quantities for ZrHfRh$_2$Bi$_2$ with those of ZrRhBi and HfRhBi. The phonon scattering rates of the QHH compound are found to be comparable to those of the HH compounds, whereas $v_g$ is significantly reduced over a broad frequency range. To quantify these results, we calculated mode average phonon group velocity ($\bar{v_g}$) and relaxation time ($\bar{\tau}$) of QHH compounds ZrHfRh$_2$Bi$_2$ and Zr$_2$RhIrBi$_2$, and HH compounds ZrRhBi, ZrIrBi, and HfRhBi using equations 
\begin{equation}
    \bar{v_g} = \sqrt{\frac{\sum_{\lambda} C_{V \lambda} v_{\lambda}^{2} \tau_{\lambda}}{\sum_{\lambda} C_{V \lambda} \tau_{\lambda}}},
\end{equation}
\begin{equation}
    \bar{\tau} = \frac{\sum_{\lambda} C_{V \lambda} v_{\lambda}^{2} \tau_{\lambda}}{\sum_{\lambda} C_{V \lambda}  v_{\lambda}^{2}},
\end{equation}
where $C_V$ is phonon specific heat and $\lambda$ is phonon mode, which is shown in TABLE \ref{tab:average}. We observe that $\bar{\tau}$ is reduced slightly in QHH compounds, however the reduction of $\bar{v_g}$ in QHH compounds is very pronounced. This reduction suppressed the overall phonon mean free path ($l_{ph}$) in the QHH compounds, as shown in FIG. \ref{fig:kappa}e for ZrHfRh$_2$Bi$_2$ and \textcolor{blue}{S15}c for Zr$_2$RhIrBi$_2$. We further calculated average Gr\"uneisen parameters ($\bar\gamma$)
\begin{equation}
    \bar{\gamma} = \frac{\sum_{\lambda} \left | \gamma_{\lambda} \right |C_{V \lambda}} {\sum_{\lambda} C_{V \lambda}},
\end{equation}
which is slightly lower for the QHH compounds. Thus, the reduction in $\kappa_p$ in QHH compounds is predominantly contributed by the suppressed $v_g$ rather than an enhancement of anharmonicity. The shorter phonon relaxation times provide an additional reduction in the $l_{ph}$, which suppresses the overall $\kappa_l$ in QHH compounds.

\subsection{Thermoelectric performance}

\begin{figure}[t] 
  \centering
  \includegraphics[width=\columnwidth]{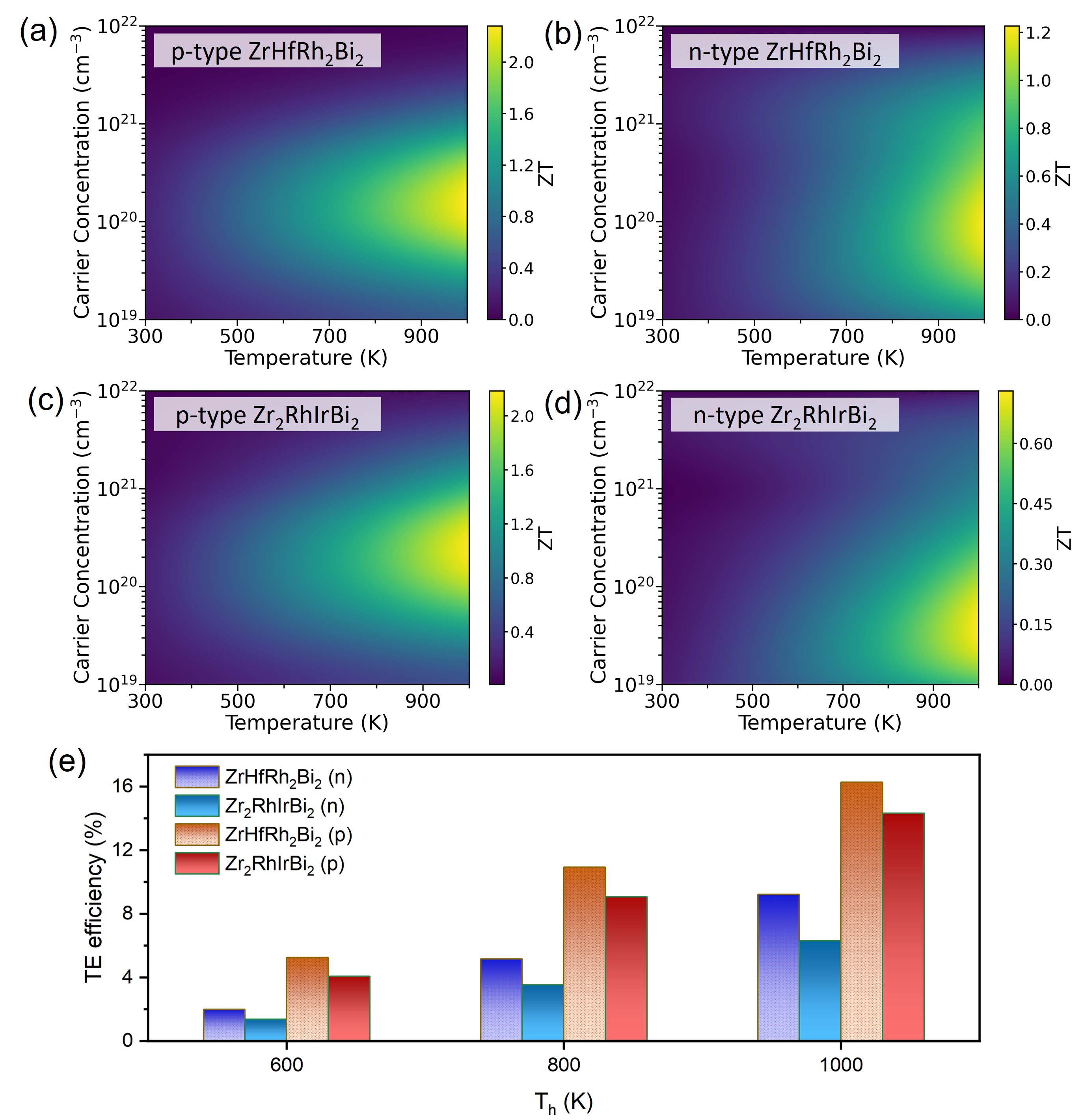}
  \caption{Thermoelectric figure-of-merit ($ZT$) of (a) p-type ZrHfRh$_2$Bi$_2$, (b) n-type ZrHfRh$_2$Bi$_2$, (c) p-type Zr$_2$RhIrBi$_2$, and (d) n-type Zr$_2$RhIrBi$_2$ as functions of temperature and charge carrier concentration. (e) Thermoelectric power conversion efficiency ($\eta$) vs heat source temperature $T_h$, where heat sink temperature $T_c=300$ K.}
  \label{fig:ZT}
\end{figure}

The high power factor and significantly suppressed lattice thermal conductivity lead to excellent thermoelectric performance in the QHH compounds. FIG. \ref{fig:ZT}a-\ref{fig:ZT}d show the variation of $ZT$ with carrier concentration and temperature for both p-type and n-type ZrHfRh$_2$Bi$_2$ and Zr$_2$RhIrBi$_2$. We obtain the highest $ZT$ values of 2.28 and 2.18 for p-type ZrHfRh$_2$Bi$_2$ and p-type Zr$_2$RhIrBi$_2$, respectively, at 1000 K, corresponding to optimal carrier concentrations of $1.55\times10^{20}$ cm$^{-3}$ and $2.44\times10^{20}$ cm$^{-3}$. In comparison, the maximum $ZT$ values for the n-type compounds are 1.22 for n-type ZrHfRh$_2$Bi$_2$ and 0.73 for n-type Zr$_2$RhIrBi$_2$, obtained at carrier concentrations of $0.95\times10^{20}$ cm$^{-3}$ and $0.73\times10^{20}$ cm$^{-3}$, respectively. The corresponding thermoelectric conversion efficiency ($\eta$) is calculated using\cite{gayner2016recent} 
\begin{equation}
\eta = \frac{T_h-T_c}{T_h} \frac{\sqrt{1+ZT_m}-1}{\sqrt{1+ZT_m}+\frac{T_c}{T_h}}
\label{eq:eta}
\end{equation}
where the first part determines Carnot's efficiency of a system operating between heat source ($T_h$) and heat sink ($T_c$). Here, $ZT_m$ is the average thermoelectric figure of merit, obtained by integrating $ZT$ throughout the temperature regime i.e. $ZT_m = \dfrac{1}{T_h-T_c}\int ZT~dT$.
Assuming a cold-side temperature ($T_c$) of 300 K and a hot-side temperature ($T_h$) of 1000 K, the maximum conversion efficiencies are $\sim$16.3\% for p-type ZrHfRh$_2$Bi$_2$ and 14.3\% for p-type Zr$_2$RhIrBi$_2$ (FIG. \ref{fig:ZT}e). These results demonstrate that the proposed quaternary alloying strategy effectively enhances the thermoelectric performance of half-Heusler compounds, yielding both high $ZT$ and excellent energy conversion efficiency.

\subsection{Effect of Grain Boundary Scattering}

\begin{figure}[t] 
  \centering
  \includegraphics[width=\columnwidth]{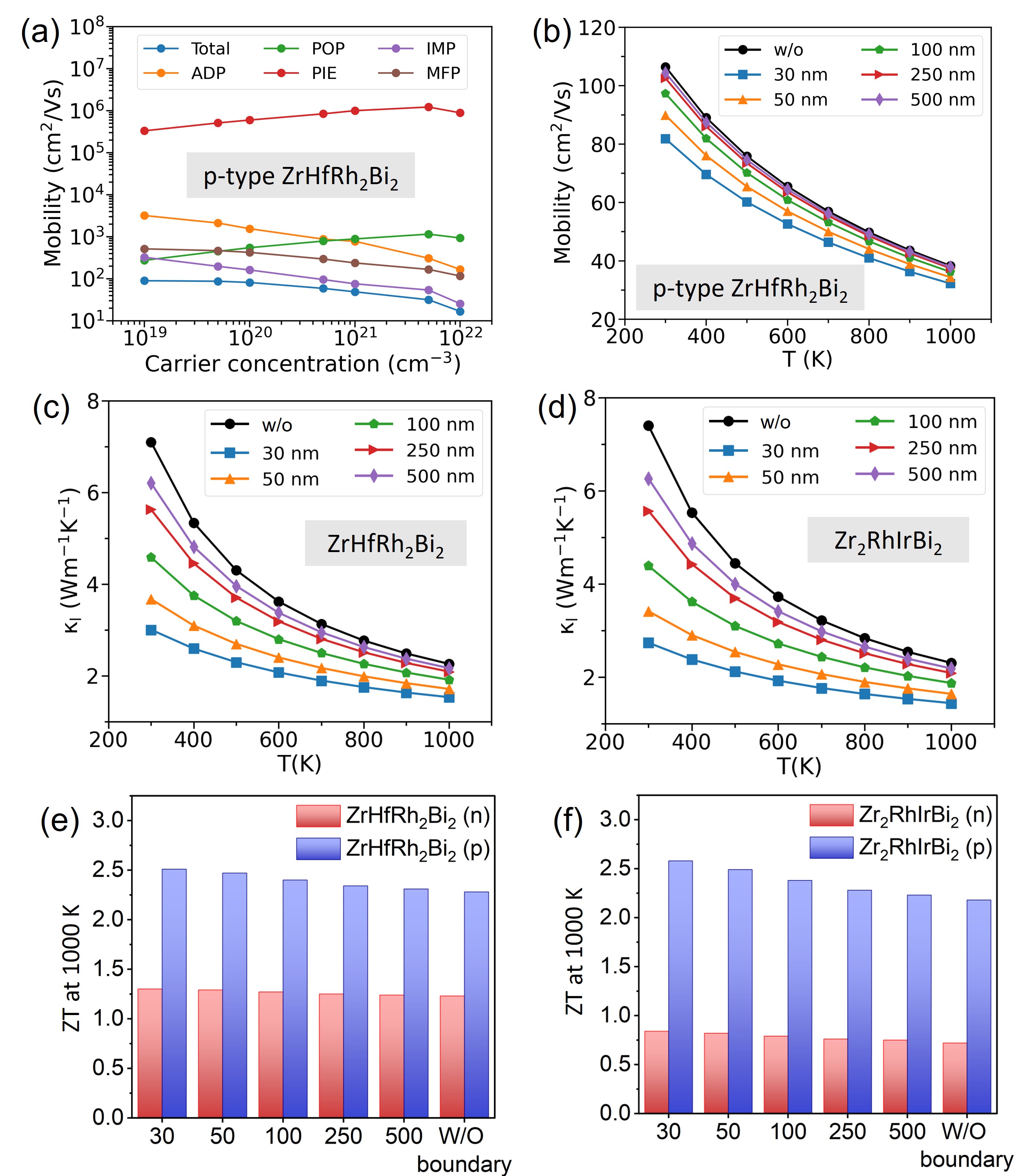}
  \caption{(a) Contributions of different scattering mechanisms to the charge-carrier mobility, including grain-boundary scattering (MFP), for p-type ZrHfRh$_2$Bi$_2$ with a grain size of 30 nm. (b) Temperature-dependent charge-carrier mobility for different grain sizes from 30 to 500 nm at a carrier concentration of $1\times10^{20}$ cm$^{-3}$ in p-type ZrHfRh$_2$Bi$_2$. Grain-size-dependent lattice thermal conductivity, $\kappa_l$, for (c) ZrHfRh$_2$Bi$_2$ and (d) Zr$_2$RhIrBi$_2$. Grain-size-dependent $ZT$ at 1000 K, for (e) ZrHfRh$_2$Bi$_2$ and (f) Zr$_2$RhIrBi$_2$.}
  \label{fig:boundary}
\end{figure}

Further, we investigated the effect of grain boundary scattering on the thermoelectric performance of the QHH compounds by considering grain sizes of 30, 50, 100, 250, and 500 nm. The grain size of polycrystalline thermoelectric materials can be effectively controlled through synthesis conditions, sintering temperature, annealing treatment, and other microstructural engineering techniques. Such control directly influences grain boundary scattering of both charge carriers and phonons, thereby affecting the overall thermoelectric performance. FIG. \ref{fig:boundary}a shows that grain boundary scattering (MFP) for a grain size of 30 nm is the second-largest contributor to the total carrier scattering in p-type ZrHfRh$_2$Bi$_2$ at 300 K after the impurity scattering (IMP). Thus, the total carrier mobility decreases from 106.4 cm$^2$ V$^{-1}$ s$^{-1}$ in the absence of grain boundary scattering to 81.9 cm$^2$ V$^{-1}$ s$^{-1}$ for a grain size of 30 nm. However, grain boundaries scatter long-wavelength heat-carrying acoustic phonons much more effectively than charge carriers, resulting in a substantially larger reduction in $\kappa_l$. As shown in FIG. \ref{fig:boundary}c and \ref{fig:boundary}d, $\kappa_l$ is reduced to $\sim$58\% and $\sim$63\% for ZrHfRh$_2$Bi$_2$ and Zr$_2$RhIrBi$_2$, respectively at 300 K. The larger reduction in $\kappa_l$ enhances the overall $ZT$ of these QHH compounds, reaching maximum values of 2.51 for p-type ZrHfRh$_2$Bi$_2$ and 2.58 for p-type Zr$_2$RhIrBi$_2$ at a grain size of 30 nm and a temperature of 1000 K (FIG. \ref{fig:boundary}e and \ref{fig:boundary}f). Overall, the combined effects of quaternary alloying and grain boundary engineering significantly improve the thermoelectric performance, making these compounds promising candidates for high-temperature thermoelectric energy conversion. 

\section{Conclusion}

In this work, we systematically investigated the thermoelectric properties of the quaternary half-Heusler compounds ZrHfRh$_2$Bi$_2$ and Zr$_2$RhIrBi$_2$. Both compounds exhibit favorable electronic transport properties with high power factors. The analysis of carrier scattering mechanisms shows that ionized impurity scattering dominates the carrier mobility, while polar optical phonon scattering becomes important at low carrier concentrations. The $\kappa_l$ was analyzed by considering both the populations($\kappa_p$) and coherence contributionsc ($\kappa_c$). Compared with their parent ternary HH compounds, quaternary alloying significantly suppresses $\kappa_p$ through reduced phonon $v_g$, while the reduced phonon interband spacing enhances the coherence contribution. Nevertheless, the overall $\kappa_l$ is substantially reduced, leading to significantly improved thermoelectric performance. We found maximum $ZT$ values of 2.28 and 2.18 are achieved for p-type ZrHfRh$_2$Bi$_2$ and Zr$_2$RhIrBi$_2$, respectively, at 1000 K.

Furthermore, grain boundary engineering has been shown to provide an additional route to enhance the thermoelectric performance of these compounds. Although grain boundary scattering slightly reduces the carrier mobility, it suppresses $\kappa_l$ more effectively by scattering long-wavelength heat-carrying acoustic phonons. As a result, the maximum $ZT$ further increases to 2.51 for p-type ZrHfRh$_2$Bi$_2$ and 2.58 for p-type Zr$_2$RhIrBi$_2$ for a grain size of 30 nm and at 1000 K temperature. This study demonstrates that the combination of quaternary alloying and grain boundary engineering is an effective strategy for realizing high-performance half-Heusler thermoelectric materials for high-temperature energy conversion applications.

\begin{acknowledgements}
S.G. sincerely acknowledges ANRF, Govt. of India for providing the National Post-doctoral Fellowship (File number: PDF/2025/001037). S.P. thanks CSIR, Government of India, for providing the SRF CSIR Fellowship. S.K.P. acknowledges ANRF JC Bose National Fellowship (File number: ANRF/SKP/4719) and SERB, Govt. of India for funding.
\end{acknowledgements}

\bibliographystyle{apsrev4-2}
\bibliography{mybib}
\end{document}